\documentclass{jfm}
\usepackage{graphicx}
\usepackage{natbib}

\ifCUPmtlplainloaded \else
  \checkfont{eurm10}
  \iffontfound
    \IfFileExists{upmath.sty}
      {\typeout{^^JFound AMS Euler Roman fonts on the system,
                   using the 'upmath' package.^^J}%
       \usepackage{upmath}}
      {\typeout{^^JFound AMS Euler Roman fonts on the system, but you
                   dont seem to have the}%
       \typeout{'upmath' package installed. JFM.cls can take advantage
                 of these fonts,^^Jif you use 'upmath' package.^^J}%
      }
  \else
  \fi
\fi

\ifCUPmtlplainloaded \else
  \checkfont{msam10}
  \iffontfound
    \IfFileExists{amssymb.sty}
      {\typeout{^^JFound AMS Symbol fonts on the system, using the
                'amssymb' package.^^J}%
       \usepackage{amssymb}%
       \let\le=\leqslant  
       \let\ge=\geqslant  
      }{}
  \fi
\fi

\ifCUPmtlplainloaded \else
  \IfFileExists{amsbsy.sty}
    {\typeout{^^JFound the 'amsbsy' package on the system, using it.^^J}%
     \usepackage{amsbsy}}
    {\providecommand\boldsymbol[1]{\mbox{\boldmath $##1$}}}
\fi

\newcommand\rme{\mathrm{e}}
\newcommand\rmi{\mathrm{i}}
\newcommand\rmd{\mathrm{d}}

\providecommand\bnabla{\boldsymbol{\nabla}}

\newcommand\bn{\hat{\mathbf{n}}}
\newcommand\bt{\hat{\mathbf{t}}}
\newcommand\ux{\hat{\mathbf{x}}}
\newcommand\uy{\hat{\mathbf{y}}}
\newcommand\bx{\mathbf{x}}
\newcommand\bu{\mathbf{u}}
\newcommand\bU{\mathbf{U}}

\newcommand\xd{\dot{x}}
\newcommand\yd{\dot{y}}

\newcommand\ydd{\ddot{y}}

\newcommand\pL{p_{\mathrm{reac.}}}
\newcommand\pT{p_{\mathrm{resis.}}}

\newcommand\Lin{\mathcal{L}}
\newcommand\fL{f_{\mathrm{reac.}}}
\newcommand\fT{f_{\mathrm{resis.}}}

\newcommand\hot{\mbox{h.o.t.}}
\newcommand\cc{\mbox{c.c.}}

\title{The origin of hysteresis in the flag instability}

\author[C. Eloy, N. Kofman and L. Schouveiler]%
{C\ls H\ls R\ls I\ls S\ls T\ls O\ls P\ls H\ls E\ns     E\ls L\ls O\ls Y$^{1,2}$%
  \thanks{Email address for correspondence: eloy@irphe.univ-mrs.fr},\ns
N\ls I\ls C\ls O\ls L\ls A\ls S\ns      K\ls O\ls F\ls M\ls A\ls N$^1$\break
\and L\ls I\ls O\ls N\ls E\ls L\ns      S\ls C\ls H\ls O\ls U\ls V\ls E\ls I\ls L\ls E\ls R$^1$}

\affiliation{$^1$IRPHE, CNRS \& Aix-Marseille Universit\'e, 49 rue Joliot-Curie, 13013 Marseille, France\\[\affilskip]
$^2$Department of Mechanical and Aerospace Engineering, University of California San Diego,\\ La Jolla, CA 92093, USA}

\date{?; revised ?; accepted ?. - To be entered by editorial office}
\begin{document}

\maketitle

\begin{abstract}
The flapping flag instability occurs when a  flexible cantilevered plate is immersed in a uniform airflow. To this day, the nonlinear aspects of this aeroelastic instability are largely unknown. In particular, experiments in the literature all report a large hysteresis loop, while the bifurcation in numerical simulations is either supercritical or subcritical with a small hysteresis loop. In this paper, this discrepancy is addressed. First weakly nonlinear stability analyses are conducted in the slender-body and two-dimensional limits, and second new experiments are performed with flat and curved plates. The discrepancy is attributed to inevitable  planeity defects of the plates in the experiments.
\end{abstract}

\section{Introduction}
\label{sec:intro}

The scientific interest in the flapping flag instability dates back to Lord \cite{Rayleigh79}.
As a side remark in his famous paper on the stability of jets, he showed that an infinite membrane placed in an airflow is always unstable.
Of course, the problem becomes  more complex when bending rigidity and finite plate dimensions are taken into account \citep[see][for recent reviews]{Paidoussis_book2,Shelley2011}

Theoretical models of this instability can be divided into two categories according to the plate aspect ratio. For large aspect ratios, a two-dimensional analysis is relevant, and
\cite{Kornecki76} were the first to show that the flow around the plate can be modelled using  unsteady airfoil theory.
When the plate aspect ratio is asymptotically small however, the aerodynamic forces can be modelled using slender-body theory \citep{Datta75,Lemaitre2005}.
The studies  in these two asymptotic limits have been recently generalised by \cite{Eloy2007} and \cite{Doare2011} who considered intermediate aspect ratios and confinement effects.

The aforementioned stability analyses are all linear though and, to our knowledge, the nonlinear dynamics has never been addressed theoretically. This is one of the goal of this paper, the other being to compare these new weakly nonlinear analyses to experiments.

Since the pioneering study of \cite{Taneda68}, different groups have performed experiments on the flag instability either for small aspect ratios \citep{Datta75,Lemaitre2005} or for moderate to large aspect ratios \citep[][among others]{Zhang2000,Tang2003,Eloy2008}.
In this latter case, a large hysteresis is always observed at threshold or, said differently, the motionless state and the flapping state coexist and are both stable over a large range of airflow velocities. To quantify the importance of this hysteresis, we will refer to the \emph{hysteresis loop}, $(U_c-U_d)/U_c$, where $[U_d\;\; U_c]$ is the velocity range of bistability. In large aspect ratio experiments, this hysteresis loop is typically of $20\%$. For aspect ratios smaller than one however, hysteresis can disappear \citep{Eloy2008}.

For small aspect ratios, and negecting the nonlinearities of the aerodynamical forces, \cite{Yadykin2001} showed that elastic and inertial nonlinearities yield no hysteresis. In the two-dimensional limit, most numerical simulations have exhibited bistability, both for inviscid flows modelled with vortex methods \citep{Alben2008,Michelin2008} or for viscous flows modelled with Navier Stokes solvers \citep{Zhu2002,Connell2007}. The only exceptions are the studies of \cite{Tang2003} who did not consider aerodynamical nonlinearities and \cite{Tang2007} who considered relatively short and heavy plates.
Even when bistability is present, the hysteresis loop is much smaller than in the experiments:
\cite{Alben2008} and \cite{Michelin2008} report an hysteresis loop of 2.5--4\% and 4.5\% while experiments in the same range of parameters exhibit loops of 20\% or more. Several effects have been advanced to account for this discrepancy: blockage and confinement effects \citep{Tang2007}, planeity defects  of the plate \citep{Tang2007,Eloy2008}, or damping effects \citep{Alben2008}.
But, until now, none of these hypotheses have been tested.

This paper is organised as follows: in~\S\,\ref{sec:setup} the experimental setup is briefly described; in~\S\,\ref{sec:model} the physical model is introduced and the weakly nonlinear analyses are carried out; in~\S\,\ref{sec:results} the experimental results are presented and compared to the theoretical predictions; and finally in~\S\,\ref{sec:discussion} these results are discussed.

\section{Experiments}
\label{sec:setup}

\begin{figure}
\centerline{	\includegraphics[scale=0.375]{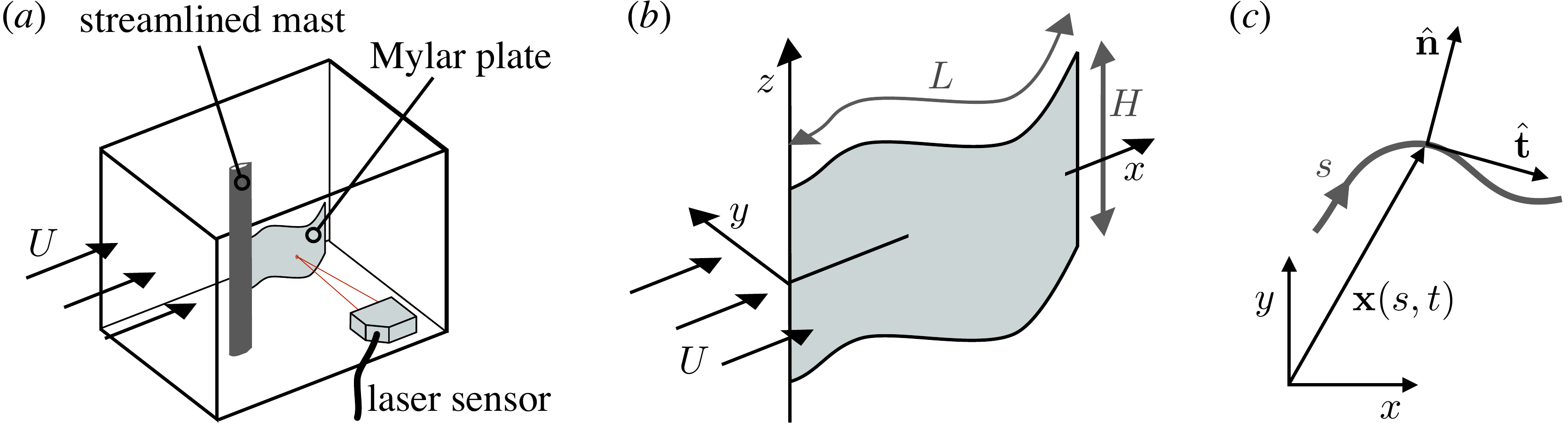}}
  \caption{(\textit{a}) Sketch of the experimental setup; (\textit{b}) definition of the coordinates and plate dimensions; and (\textit{c}) view of the one dimensional plate motion $\bx(s,t)=(x,y)$.}
  \label{fig:sketch}
\end{figure}

The experimental set-up is illustrated in~figure~\ref{fig:sketch}\textit{a}. Experiments are performed in a horizontal
low-turbulence wind tunnel of $80\times 80\;$cm$^2$ cross section. The flexible rectangular plates are cut from Mylar sheets whose physical characteristics are given in table~\ref{tab:parameters}. The bending rigidity $D$ has been measured through deflection  tests under gravity, and
the fluid and material damping coefficients $\nu$ and $\mu$ have been evaluated by measuring the damping of clamped plates in air at rest.

The same protocol has been followed in all experiments and will be briefly described.
Once the plate is clamped into the mast, the flow velocity is gradually increased. At small velocities the plate is motionless. Eventually, for a critical flow velocity $U_c$, the plate starts to flutter with a large amplitude and a well-defined frequency. When the flow velocity is decreased, the plate returns to its stable state again at a different critical velocity $U_d \le U_c$, thus leading to a hysteretic cycle.

The motion of the plate is recorded with a CCD laser displacement sensor of spatial and frequencies accuracies of 1\,$\mu$m and $10\,000\,$Hz. In the present study, the deflection was measured at the plate center. At this point and near threshold the deflection is always harmonic in time, and the amplitude $A$ will refer to the peak amplitude of the deflection.

\begin{table}
  \begin{center}
\def~{\hphantom{0}}
  \begin{tabular}{@{}lccl@{}}
      Parameter  				& Symbol	& Value  		& Unit \\[3pt]
      Plate length				& $L$		& 3--30			& cm\\
      Plate height				& $H$		& 1--20			& cm\\
      Plate surface density	& $m$		& 0.25			& kg\,m$^{-2}$\\
      Plate bending rigidity	& $D$		& $3.0\times 10^{-3}$		& N\,m\\
      Fluid damping		  		&$\nu$		& 0.044 		& kg\,m$^{-2}$\,s$^{-1}$\\
      Internal damping  		&$\mu$		& $1.7\times 10^{-6}$ 		& kg\,m$^{2}$\,s$^{-1}$\\
      Wind velocity				& $U$		& 0--65			& m\,s$^{-1}$ \\
      Air density				& $\rho$	& 1.20			& kg\,m$^{-3}$	\\
  \end{tabular}
  \caption{Characteristics of the experiments.}
  \label{tab:parameters}
  \end{center}
\end{table}

\section{Modelling}
\label{sec:model}
Consider a flexible plate of length $L$ and span $H$ clamped into a mast parallel to the vertical axis $0z$ and immersed in a uniform flow of velocity $U$ (figure~\ref{fig:sketch}\textit{b}). Assuming the motion is independent of $z$, this plate obeys the Euler-Bernoulli beam equation
\begin{equation}\label{eq:motion}
m \partial_{t}^2 \bx + D  \partial_{s}^4 \bx -\partial_{s}(\langle T \rangle\partial_{s}\bx ) +
\nu\partial_{t} \bx + \mu\partial_{t}\partial_{s}^4 \bx +\langle p \rangle \bn =0,
\label{eq:Euler}
\end{equation}
with $s$ the curvilinear coordinate, $\bx(s,t)=(x,y)$ the plate position,  $m$ its mass per unit area, $D$ its bending rigidity, $T$ the generalised tension in the plate that enforces inextensibility (i.e. $\| \partial_{s}\bx \|^2=1$), and the chevrons $\langle . \rangle$ denote the average along the span. The terms proportional to  $\nu$ and $\mu$ in (\ref{eq:Euler}) model the dissipation due to the fluid viscosity and the Kelvin-Voigt structural damping respectively \citep{Paidoussis_book2}.
The last term in~(\ref{eq:Euler}) is due to the pressure jump across the plate, $p(s,z,t)$, and $\bn$ is the unit vector normal to the plate (figure~\ref{fig:sketch}\textit{c}).
In addition the deflection $y(s,t)$ satisfies clamped-free boundary conditions: $y=\partial_{s}y=0$ in $s=0$, and $\partial^2_{s}y=\partial^3_{s}y=T=0$ in $s=L$.

The Reynolds number being large, we will further assume that the flow is potential such that the total flow velocity is given by $\bu=\bU+\bnabla\phi$, with $\bU$ the imposed airflow and $\phi$ the perturbation potential. The potential $\phi$ can be found by solving a Laplace problem with Neumann boundary conditions ensuring the impermeability of the plate
\begin{equation}
\Delta \phi=0, \quad \mbox{with} \quad \bnabla \phi\cdot\bn = (\partial_{t} \bx - \bU) \cdot\bn \quad \mbox{on } S,
\label{eq:Laplace}
\end{equation}
where $S$ is the moving plate surface. From $\phi$  the pressure jump can be calculated using the unsteady Bernoulli equation.

Using $L$ and $L/U$ as characteristic length and time, the control parameters are reduced to three dimensionless numbers noted with asterisks: the reduced velocity $U^*$, the mass ratio $M^*$, and the aspect ratio $H^*$
\refstepcounter{equation}
$$U^*= UL\sqrt{m/D}, \quad M^*=\rho L/m, \quad H^*=H/L,
\eqno{(\theequation{\mathit{a}\mbox{--}\mathit{c}})}
$$
and the dissipative properties are characterised by two coefficients
\refstepcounter{equation}
$$
\nu^*= \nu\rho^{-2}m^{\frac{3}{2}}D^{-\frac{1}{2}}, \quad \mu^*= \mu\rho^{2}m^{-\frac{5}{2}}D^{-\frac{1}{2}},
\eqno{(\theequation{\mathit{a},\mathit{b}})}
$$
where $\rho$ is the fluid density.
With this non-dimensionalisation the damping coefficients depend neither on the plate dimensions $H$ and $L$ nor on the flow velocity $U$ and thus are constant in the present study : $\nu^*=0.071\pm0.001$ and $\mu^*=0.0014\pm0.0003$ (with 95\% confidence intervals).

Equations~(\ref{eq:Euler}--\ref{eq:Laplace}) describe the nonlinear, fully coupled, fluid-structure interaction. We will now consider three approximations of these equations: the linear limit to address the instability threshold, and the weakly nonlinear limit both for an elongated plate and for an infinite plate, to study the nature of the bifurcation.

\subsection{Linear model}
\label{sec:linear.model}
Assuming small lateral deflection ($y\ll L$ and $\partial_x y\ll 1$), the system (\ref{eq:Euler}--\ref{eq:Laplace}) is linearised
\begin{subeqnarray}\label{eq:LaplaceL}
m \partial_{t}^2 y + D  \partial_{x}^4 y +
\nu\partial_{t}y + \mu\partial_{t}\partial_{x}^4y +\langle p\rangle  & = &  0, \\
\Delta \phi=0, \quad \mbox{with} \quad \partial_{y} \phi  =  \partial_{t} y + U\partial_x y\quad \mbox{for } y &= &0 \mbox{ and } 0<x<L.
\end{subeqnarray}
This  system is solved by the same method as in our previous papers \cite[][]{Eloy2007,Eloy2008}, except that  dissipative terms are retained in the analysis. The main steps of this linear stability analysis are as follow. First, a complex angular frequency $\omega$ is assumed and the deflection $y$ is expanded on Galerkin modes that satisfy the clamped-free boundary conditions. Second,  the Laplace problem (\ref{eq:LaplaceL}\textit{b}) for the perturbation potential is solved for each Galerkin mode in three dimensions, and the associated average pressure jump $\langle p\rangle$ is calculated. Finally, the eigenvalue problem obtained from the equation of motion (\ref{eq:LaplaceL}\textit{a}) is solved to obtain the global modes and their complex frequencies. If one of these complex frequencies has a negative imaginary part, the plate is unstable.

This linear stability analysis allows to predict the critical velocity $U^*_{c}$, above which the system is unstable, as a function of the dimensionless parameters $M^*$, $H^*$, $\nu^*$, $\mu^*$.

\subsection{Slender-body nonlinear model }
\label{sec:SBnonlinear.model}

In the slender body limit (i.e. $H\ll L$), the aerodynamic force on the plate can be decomposed into two terms: a reactive force originating from the added mass of air accelerated when the plate moves, and a resistive force modelling the drag on the plate due to crossflow.
As a result, the average pressure jump is decomposed as $\langle p \rangle= \pL + \pT$, where $\pL$ is the reactive part and has been calculated by \cite{Lighthill1971} in the context of fish locomotion, and $\pT$ is the resistive part
\refstepcounter{equation}
$$
\pL = M\left( \dot w - (uw)' + \textstyle{\frac{1}{2}}w^2 \kappa\right), \quad
\pT = \textstyle{\frac{1}{2}} \rho C_d |w|w,\label{eq:pT}
\eqno{(\theequation{\mathit{a},\mathit{b}})}
$$
where dots and primes denote differentiation with respect to $t$ and $s$, $u$ and $w$ are the longitudinal and normal component of the plate velocity relative to the uniform airflow (such that $\dot\bx -\bU =u \bt + w \bn$)
\refstepcounter{equation}
$$
u=-Ux'+\xd x'+\yd y', \quad w=Uy'-\xd y'+\yd x',\label{eq:uw}
\eqno{(\theequation{\mathit{a},\mathit{b}})}
$$
$M=\pi\rho H/4$ is the added mass of air, $\kappa=y''/(1-{y'}^2)^{1/2}$ is the plate curvature and $C_d$ is a drag coefficient taken to be $C_d=1.8$ for a plate \cite[see][]{Buchak2010}.

Inserting (\ref{eq:pT}--\ref{eq:uw}) into (\ref{eq:motion}) and projecting onto $\ux$ and $\uy$ give two coupled dynamical equations for $x(s,t)$ and $y(s,t)$. Following \cite{Yadykin2001}, these equations are decoupled by first using the $x$-projection to eliminate the average tension $\langle T\rangle$. Then $x(s,t)$ and its derivatives are eliminated by using the inextensibility condition.
Finally the terms of order larger than $y^3$ are discarded yielding a weakly nonlinear dynamical equation for $y(s,t)$ that can be expressed as
\begin{equation}\label{eq:WNLformal}
\Lin (y) + mf_m(y) + Df_D(y) + M\fL(y) + \rho C_d\fT(y)=0,
\end{equation}
where $\Lin$ is a linear differential operator on $y$
\begin{equation}
\Lin (y) = m \ydd + D y'''' + M\left(U^2y''+2U\yd'+\ydd\right),
\end{equation}
and the other terms are $O(y^3)$ nonlinear terms
\begin{subeqnarray}\label{eq:fs}
f_m(y) &=& y'\int_0^s \left(\yd'^2+y'\ydd'\right)\rmd s -
			y''\int_s^L\int_0^s \left(\yd'^2+y'\ydd'\right)\rmd s\rmd s,\\
f_D(y) &=& y'''' {y'}^2 +4\,y' y'' y'''+{y''}^3,\\
\fL(y) &=& -\frac{1}{2} U^2y''{y'}^2 + U(\yd'{y'}^2 -3y''y'\yd) - 2\yd'y'\yd -\frac{1}{2}y''\yd^2 +  	y'\int_0^s \left(\yd'^2+y'\ydd'\right)\rmd s + \nonumber \\
	   & &	2(Uy''+\yd')\int_0^s \yd'y'\rmd s - y'' \int_s^L y'\left(U^2y''+2U\yd'+\ydd\right)\rmd s,\\
\fT(y) &=& \frac{1}{2} |Uy'+\yd | (Uy'+\yd),
\end{subeqnarray}
except $\fT$ which is of order $y^2$. For the sake of brevity dissipative terms have been omitted in this analysis (we have checked that they were indeed negligible).

The deflection $y$ is now assumed to be of the form
\begin{equation}\label{eq:y}
y(s,t)=(A h_0(s) + h_1(s)) \rme^{\rmi \omega t} +  \cc
\end{equation}
where $A$ is a small complex amplitude ($|A|\ll L$), $h_0$ is the solution of the linear problem $\Lin (h_0 \rme^{\rmi \omega t})=0$ with clamped-free boundary conditions and the proper normalisation to allow comparison with experiments (i.e. $h_0(L/2)=1/2$, such that $y(L/2,t)= A\cos \omega t$), $h_1$ gathers the terms of order greater than $A$, and `$\cc$' stands for `complex conjugate'.

Using standard methods of perturbation theory, the weakly nonlinear amplitude equation for $A$ is found. It consists in finding and solving the linear adjoint problem, inserting the decomposition (\ref{eq:y}) into (\ref{eq:WNLformal}), and forming the scalar product with the adjoint solution. It yields the following amplitude equation given in dimensionless form for $A^*=A/L$
\begin{eqnarray}\label{eq:amplitudeSB}
-{\omega^*}^2 A^*+ \rmi {\omega^*} U^* c_1 A^*  +
	&&\left({U^*}^2 c_2  + c_3\right)A^*  +  ({U^*}^2c_4 + \rmi {\omega^*} U^*c_5) |A^*| A^* + \nonumber  \\
& &	\quad\left({U^*}^2c_6+ \rmi {\omega^*} U^*c_7 -{\omega^*}^2 c_8 +c_9\right) |A^*|^2 A^*=0,
\end{eqnarray}
where $\omega^*=\omega L^2\sqrt{mD}$, \textmd{and the complex coefficients $c_i$ depend on the control parameters $M^*$ and $H^*$. 
Once the $c_i$'s are calculated, the solutions of the second degree equation (\ref{eq:amplitudeSB}) for ${\omega^*}$ can be obtained.} Then, values of $|A^*|$ such that the imaginary part of ${\omega^*}$ is zero give the saturated amplitude $A^*(U^*)$.

The first nonlinear effect comes from the resistive force (\ref{eq:fs}\,\textit{d}) and gives the term proportional to $|A^*| A^*$ in (\ref{eq:amplitudeSB}). As a result, $A^*$ is proportional to $(U^*-U^*_c)$ close to threshold (see figure~\ref{fig:AvsU}\textit{a}). This behaviour has to be contrasted with the usual pitchfork bifurcation for which   $A^*$ is proportional to $(U-U^*_c)^{1/2}$. In any case, the bifurcation is not subcritical, and one therefore expect no hysteresis in the slender body limit.

\subsection{Two-dimensional nonlinear model}
\label{sec:2Dnonlinear.model}

For plates of large aspect ratios ($H\gg L$), substantial assumptions have to be made to solve the weakly nonlinear problem analytically.
First the plate span and length are assumed infinite. The problem then becomes two-dimensional, and it is further assumed that the deflection is a propagating wave of amplitude $A$
\begin{equation}\label{eq:Ainfini}
y=A \cos (kx -\omega t).
\end{equation}
To calculate the flow around the plate, the perturbation potential is  expanded in powers of $A$ (i.e. $\phi=A\phi_1+A^2\phi_2+A^3\phi_3+\cdots$), \textmd{similarly to the methods used for the weakly nonlinear analysis of the Rayleigh--Taylor instability \citep{Nayfeh1969}.} Inserting this expansion into (\ref{eq:Laplace}) and using a Taylor expansion to evaluate $\phi$ on the plate, the Laplace problem is solved for the first three orders yielding
\begin{subeqnarray}
\phi_1^\pm & = & \mp 								(V - U) \sin(kx - \omega t) 	\, \rme^{\mp ky}, \\
\phi_2^\pm & = &  -   \textstyle{\frac{1}{2}}	k	(V - U) \sin(2kx - 2\omega t) 	\, \rme^{\mp 2ky},\\
\phi_3^\pm & = &  \mp \textstyle{\frac{1}{8}} k^2	(V - U) \sin{(kx - \omega t)} 	\, \rme^{\mp ky} + \hot,
\end{subeqnarray}
where $V=\omega/k$ is the wave speed of the deformation, the superscript $\pm$ corresponds to the upper and lower parts of the flow and `$\hot$' stands for `higher order terms'.

The  pressure field is deduced from the potential $\phi$ using the unsteady Bernoulli equation.
The $y$-component of the pressure force $F_P$ is then calculated by evaluating the pressure jump using a Taylor expansion  and projecting it onto the vertical axis.
Keeping only the first harmonics and the terms up to order $A^3$ yields
\begin{equation}
\label{eq:NLpressure}
F_P = F_1\left(1 - \textstyle{\frac{5}{8}} \, A^2 k^2\right), \quad \mbox{with} \quad
F_1 = 2 \rho (V-U)^2  Ak \cos (kx -\omega t).
\end{equation}

The nonlinearities originating from the inertial and elastic terms can be evaluated by using the expressions (\ref{eq:fs}\,\textit{a,b}). Note that derivatives with respect to $s$ have to be transformed into derivatives with respect to $x$ using the chain rule. Note also that, when calculating (\ref{eq:fs}\,\textit{a}), boundary terms arising from the integrations are neglected because $y(x,t)$, as given by (\ref{eq:Ainfini}), do not satisfy the clamped-free boundary condition.

Neglecting the dissipative terms, \textmd{the different nonlinearities can now be gathered to obtain a weakly nonlinear dispersion relation in dimensionless form} (with $A^*=A/L$ and $k^*=kL$)
\begin{equation}\label{eq:amplitude2D}
-{\omega^*}^2 \left(1+{\frac{1}{8}} {A^*}^2{k^*}^2\right) + {k^*}^4\left(1-{\frac{1}{2}} {A^*}^2{k^*}^2\right) -2 \frac{M^*}{|k^*|} ({\omega^*}-U^*k)^2 \left(1 - {\frac{5}{8}} \, {A^*}^2{k^*}^2\right) =0,
\end{equation}
\textmd{where the first term corresponds to inertia, the second to the bending force, and the last to the pressure force.}
Solving this second degree equation for ${\omega^*}$ and finding the values of $A^*$ for which the imaginary part of ${\omega^*}$ is zero gives the saturated amplitude $A^*(U^*)$.

When the mass ratio and the wavenumber are the same as in the experiments ($M^*=0.6$, $k^*\approx 3\pi/2$), the weakly nonlinear dispersion relation  (\ref{eq:amplitude2D}) predicts a subcritical bifurcation, with a hysteresis loop of approximately $0.1\%$ (see figure~\ref{fig:AvsU}\textit{b}). The bifurcation is always subcritical, except for small values of the mass ratio, $M^*<k^*/10$, for which it is supercritical. This prediction should be considered with caution because several approximations have been made to obtain the dispersion relation  (\ref{eq:amplitude2D}): in particular, the plate has been assumed infinite to calculate the pressure forces and we know from the linear stability analysis that finite size effects can be important.

Yet, a comparison with the inviscid numerical simulations found in the literature give good qualitative agreement.
For $M^*=0.2$, $k^*=3\pi/2$, the bifurcation is found to be supercritical (since $M^*<k^*/10$) as in \cite{Tang2007}; for $M^*=3.3$, $k^*=5\pi/2$, the hysteresis loop is equal to 5.2\%  which compares well with the $3.4\%$ reported by \cite{Alben2008}; finally for $M^*=10$, $k^*=7\pi/2$, the hysteresis loop is found to be 12.8\% while \cite{Michelin2008} found $4.5\%$.

\section{Results}
\label{sec:results}

\begin{figure}
\centerline{\includegraphics[scale=0.37]{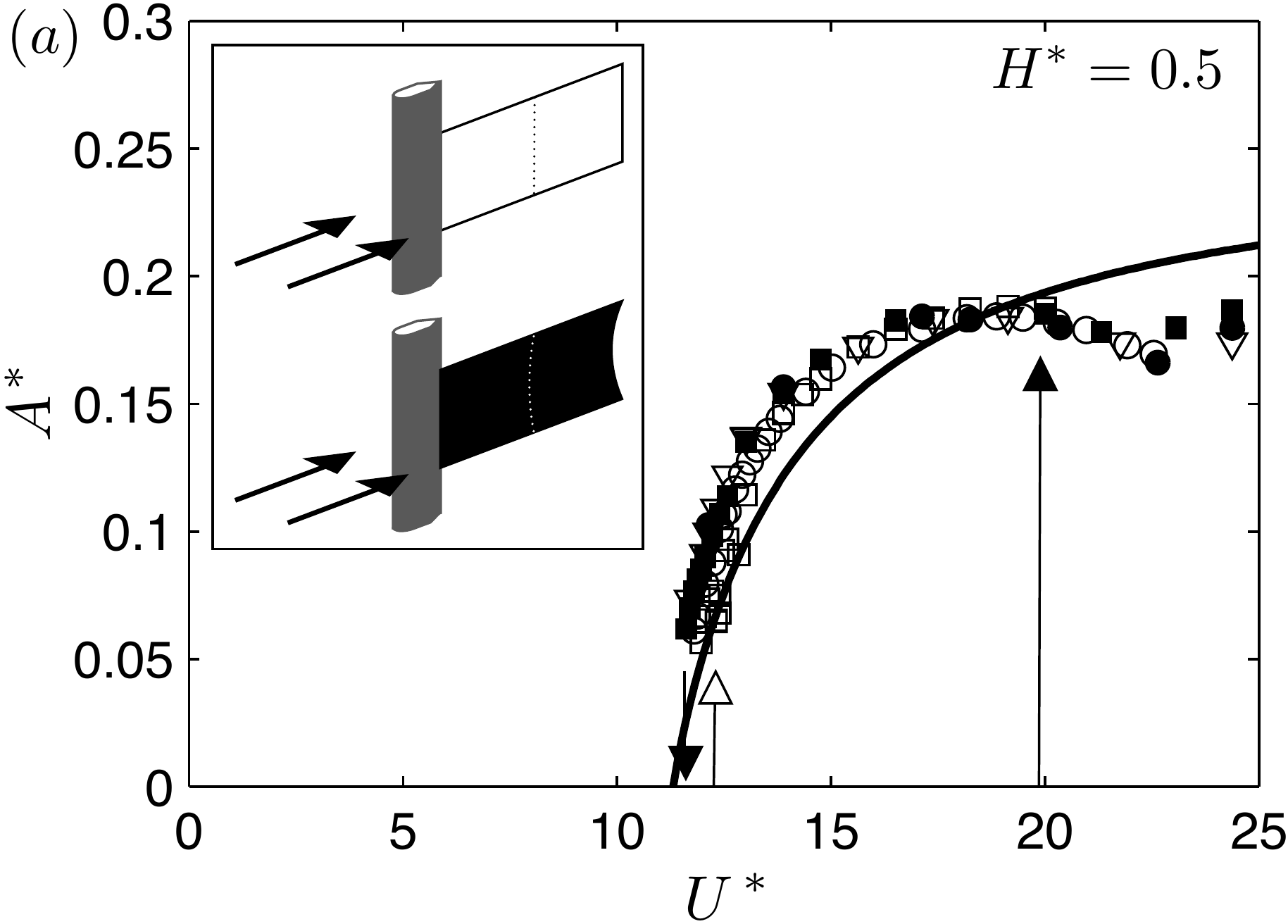}\includegraphics[scale=0.37]{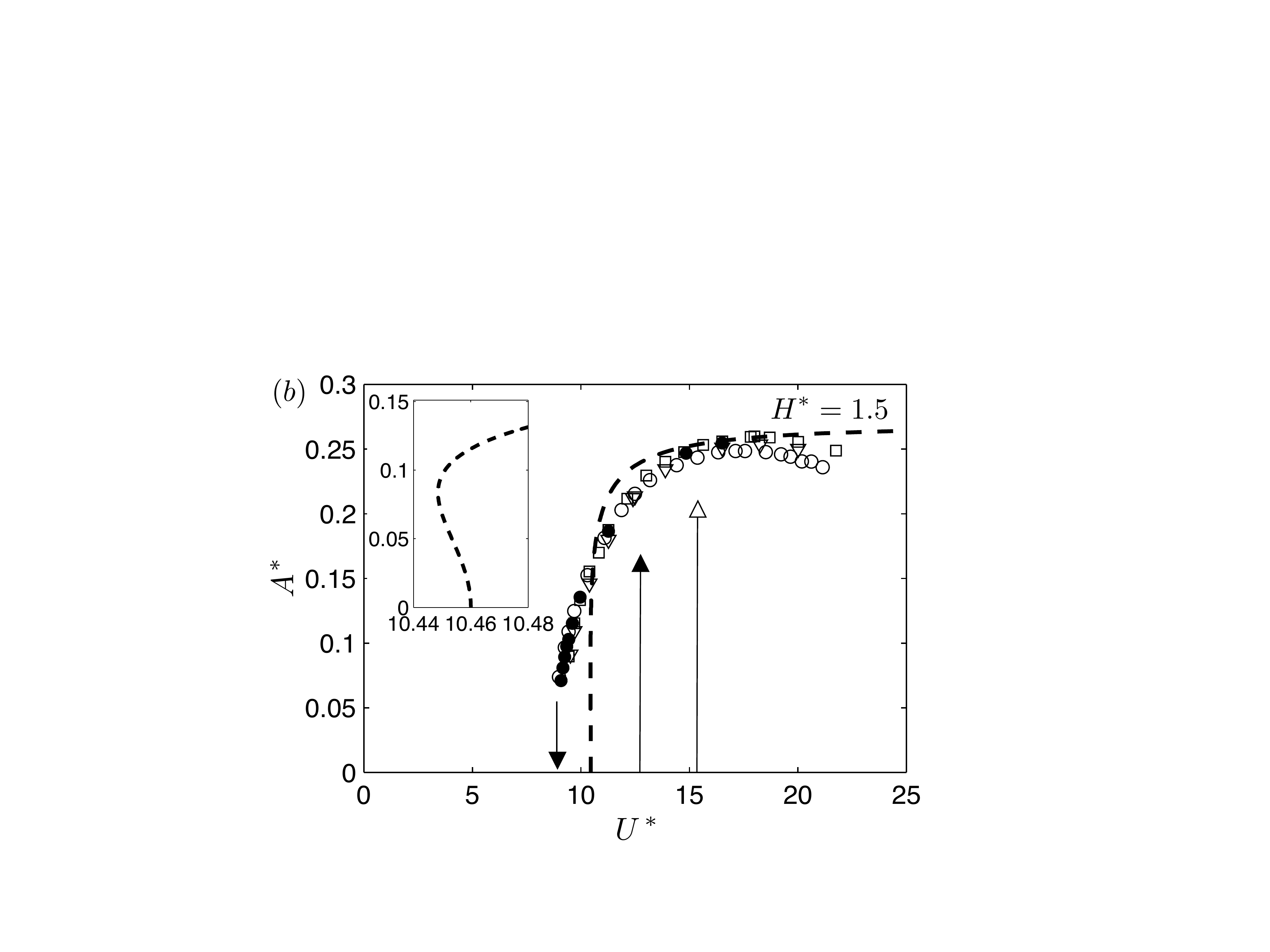}}
\centerline{\includegraphics[scale=0.37]{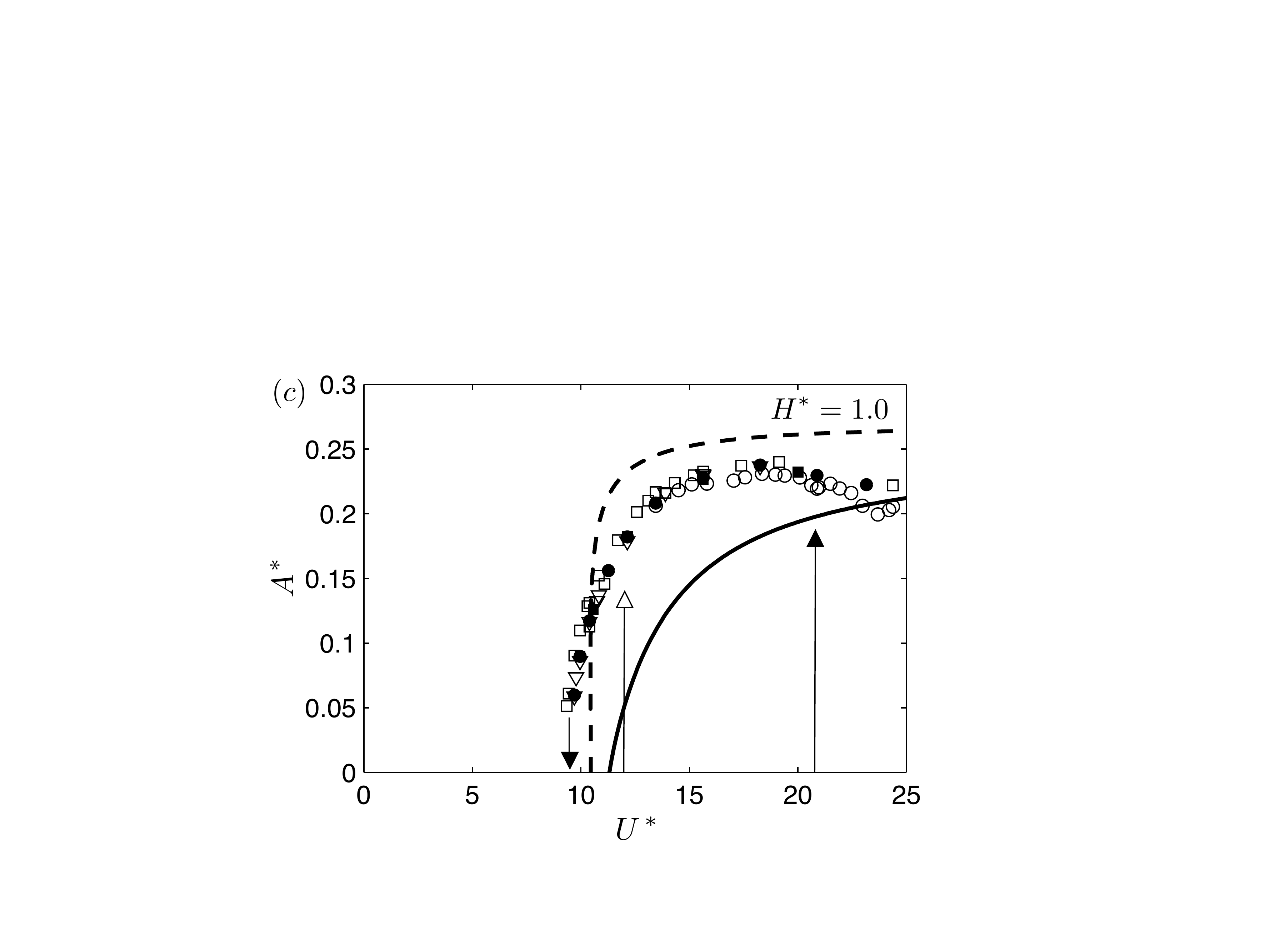}\includegraphics[scale=0.37]{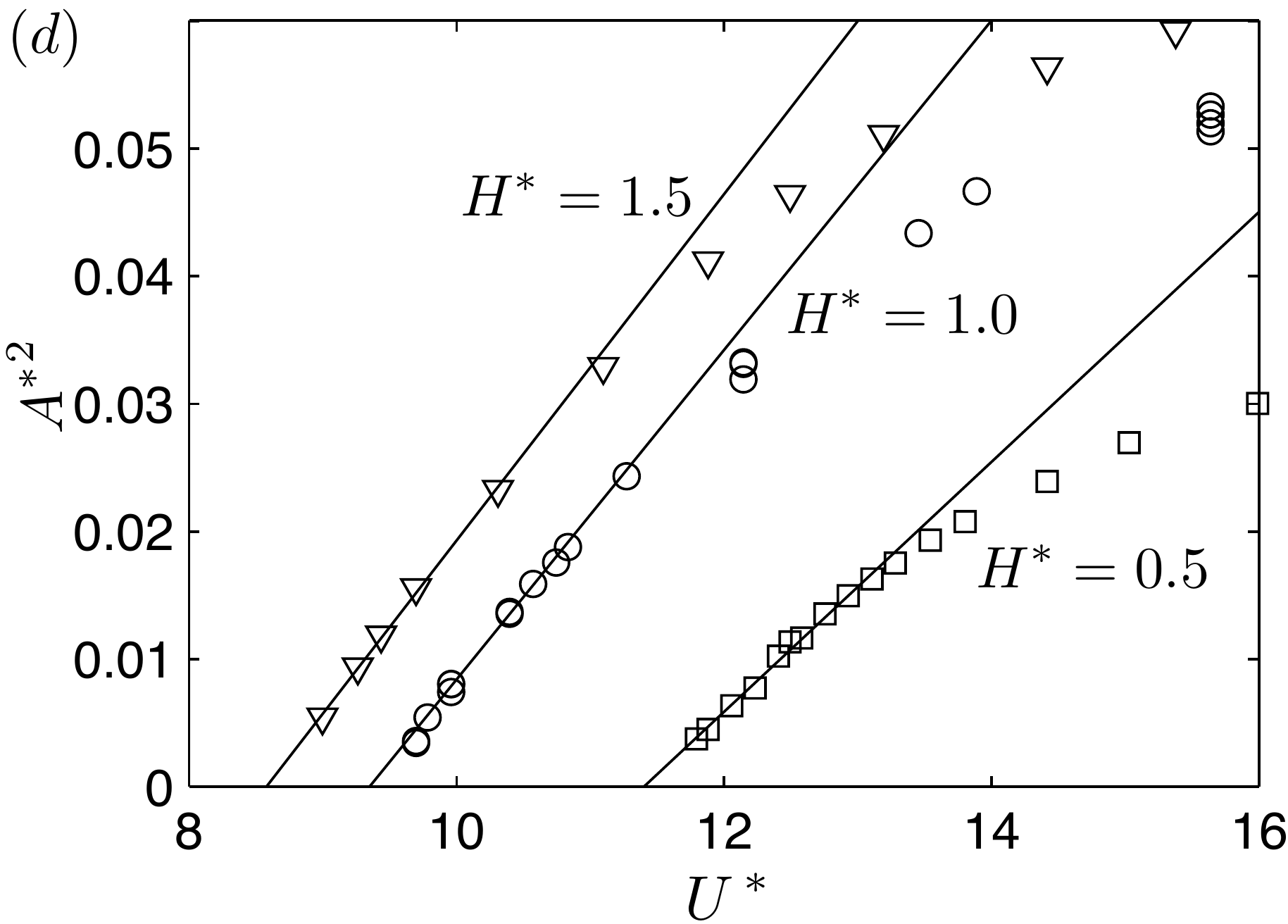}}
  \caption{Flutter amplitude as a function of the velocity $U^*$ for a mass ratio $M^*=0.6$ and aspect ratios: $H^*=0.5$ (\textit{a}), $H^*=1.5$ (\textit{b}), and  $H^*=1.0$ (\textit{c}).  Open and filled symbols correspond to experiments with flat plates and curved plates (see the inset in (\textit{a})). The corresponding average thresholds $U^*_d$ and $U^*_c$ are represented with the same convention (exact values are given in table~\ref{tab:hysteresis}). The lines in (\textit{a--c}) represent the predictions of the weakly nonlinear analyses in the slender-body (--) and two-dimensional limits (- -), as calculated from (\ref{eq:amplitudeSB}) and (\ref{eq:amplitude2D}). Finally the square of the amplitude is plotted in (\textit{d}) for three experiments with different aspect ratios as labelled (lines are  best linear fits near threshold).}
\label{fig:AvsU}
\end{figure}

To examine the dynamics near threshold, two sets of experiments have been carried out, either with \emph{flat plates}, or \emph{curved plates} (see the inset of figure~\ref{fig:AvsU}\textit{a}).
Flat plates were cut from unused Mylar sheets and clamped into the mast, whereas curved plates were first heated on a curved surface for 10\,s. In this latter case, an intrinsic curvature of the order of $0.1/L$ in the vertical direction subsisted (this curvature is exaggerated in the inset of figure~\ref{fig:AvsU}\textit{a}). 
For both flat and curved plates, each experiments was reproduced five to ten times to assess the repeatability and to extract statistics.

Experimental values of the flutter amplitude $A^*$, measured for $M^*=0.6$ and three different values of $H^*$, are plotted in figure~\ref{fig:AvsU}.
\textmd{It first shows that the deflection $A^*(U^*)$ does not depend on the intrinsic curvature of the plate and that the repeatability on this measurement is excellent.}
These plots also show that the weakly nonlinear analyses developed in~\S\,\ref{sec:model} give good predictions of the saturated flutter amplitude. In the elongated plate limit (figure~\ref{fig:AvsU}\textit{a}), the analysis predicts a supercritical bifurcation with $A^*$ proportional to $(U^*-U^*_c)$ near threshold. However, as illustrated in figure~\ref{fig:AvsU}\textit{d}, $A^*$ can always be reasonably fitted as $A^*\sim(U^*-U^*_c)^{1/2}$ for all aspect ratios, which indicates that the resistive force modelled by (\ref{eq:fs}\,\textit{d}) may not be valid for small flutter amplitudes.
Since this drag force is caused by vorticity shed from the top and bottom edges of the plate, it probably means that this shedding occurs only when the flutter amplitude is sufficiently large.

Experimental data for $H^*=1.5$ are plotted in figure~\ref{fig:AvsU}\textit{b} together with results of the two-dimensional analysis. In the inset of the figure, the scale of $U^*$ has been expanded to emphasize the subcritical nature of the bifurcation. This subcriticality does not appear on experimental data (see figure~\ref{fig:AvsU}\textit{d}), but this could be due to the  moderate value of the aspect ratio.
For $H^*=1.0$ (figure~\ref{fig:AvsU}\textit{c}), the measured flutter amplitudes lie between the predictions of the slender-body and two-dimensional approximations.

\textmd{Statistics on the the critical velocities have also been performed and the results are reproduced in table~\ref{tab:hysteresis} (see also the arrows in figure~\ref{fig:AvsU} indicating the mean values of $U^*_c$ and $U^*_d$).
These statistics show that $U^*_d$ exhibits a good repeatability and has indistinguishable values whether the plate is curved or flat. 
On the contrary, the mean value of $U^*_c$ changed drastically between flat and curved plates. In addition, the standard deviation for $U^*_c$ is consistently larger than for $U^*_d$, showing a  relatively poor repeatability.}

\textmd{These results show that $U^*_c$ is extremely sensitive to curvature. 
As it was shown by \cite{Peake1997}, the plate curvature can be taken into account in the linear regime by adding a spring foundation term to the equation of motion (\ref{eq:motion}), with spring rate per unit surface, $Eh\kappa^2$, with $E=1.5\,$GPa the plate Young's modulus, $h=280\,\mu$m the plate thickness, and $\kappa\approx 0.1\,$m$^{-1}$ the curvature. 
In dimensionless terms, this corresponds to adding a constant term of order $\delta^2/h^2$ to the dispersion relation (\ref{eq:amplitude2D}), where $\delta$ is the typical deflection due to curvature. For the curved plates  used here, this dimensionless spring rate is approximately 200 and leads to a 20\% increase of the instability threshold, in qualitative agreement with the experimental observations 
 \citep[this term is only an approximation because nonlinearities should be considered when $\delta\gtrsim 0(h)$,][]{Audoly2010}.
As soon as planeity defects are of the order of the plate thickness, and this is almost impossible to avoid in practice, one then expects a stiffening effect and a delayed instability. 
}

\textmd{When the plate starts to flutter however, one expects this curvature to be `ironed out', as originally hypothesised by \cite{Tang2007}.
This ironing out is due to the prohibitive cost of having a non-zero Gauss curvature that would introduce stretching energy in addition to bending energy.}

\textmd{In brief, we propose the following scenario. 
Because of curvature or planeity defects, the plate is stiffen and the instability is delayed. 
Once the plate flutters however, these defects are ironed out and the decreasing threshold $U^*_d$ is the one predicted by the analyses. 
This scenario is coherent with all the experimental observations. It explains why $A^*$ and $U^*_d$ are identical when curvature is added, why $U^*_c$ has poor repeatability and can be increased by curvature. It also explains why the hysteresis loop tends to increase with the aspect ratio for flat plates (curvature defects should also have a zero Gauss curvature and thus are more likely to be in the vertical direction when $H^*\ge 1$).
Note that for the largest aspect ratio, $H^*=1.5$, the curved plates have a lower mean threshold $U^*_c$ than flat plates, but this is not statistically significant because of insufficient data. }

\textmd{The present stiffening scenario might not be the only cause of hysteresis: 
in particular, it cannot explain the bistability observed in the soap film experiments of \cite{Zhang2000}.
However, the alternative explanations found in the literature are ruled out by the present set of experiments: both blockage and damping effects would not lead to poor repeatability of $U^*_c$ and a drastic increase of $U^*_c$ when curvature is introduced.
}

\begin{table}
  \begin{center}
\def~{\hphantom{0}}
  \begin{tabular}{@{\extracolsep{10pt}}l cc cc cc@{}}
   & \multicolumn{2}{c}{$H^*=0.5$}	& \multicolumn{2}{c}{$H^*=1.0$}	& \multicolumn{2}{c}{$H^*=1.5$}\\
 \cline{2-3}\cline{4-5}\cline{6-7} \noalign{\smallskip}
				& $U^*_d$	& $U^*_c$	& $U^*_d$	& $U^*_c$	& $U^*_d$	& $U^*_c$	\\[3pt]
Flat plates		& $11.7\pm 0.2$	& $12.3\pm 0.3$	& $9.4\pm 0.1$	&$12.0\pm 1.8$	&$9.0\pm 0.4$	&$15.4\pm 4.9$	\\
Curved plates	& $11.5\pm 0.1$	& $19.9\pm 3.7$	& $9.5\pm 0.1$	&$20.8\pm 2.9$	&$8.7\pm 0.4$	&$12.8\pm 1.5$	
  \end{tabular}
  \caption{Statistics of the measured critical velocities $U_d^*$ and $U_c^*$ for $M^*=0.6$ and three aspect ratios $H^*=0.5$, $1.0$, and $1.5$. For each case, the first figure is the mean and the second the standard deviation measured on 5 to 10 similar experiments. }
  \label{tab:hysteresis}
  \end{center}
\end{table}

Finally, it should be noted that no amplitude smaller than $A^*\approx 0.06$ could be measured, for all aspect ratios and despite the care taken.
It is thus difficult to give a definite answer on the nature of the bifurcation. Nevertheless, the amplitude clearly behaves as $A^*\sim(U^*-U^*_c)^{1/2}$ near threshold (figure~\ref{fig:AvsU}\textit{d}) and we have shown that the hysteresis is mainly caused by planeity defects. Without these defects, we would thus expect a supercritical bifurcation, in agreement with the theory (the bifurcation could become subcritical for larger $H^*$ or larger $M^*$, but then, plates may sag under their own weight).

\begin{figure}
\centerline{\includegraphics[scale=0.37]{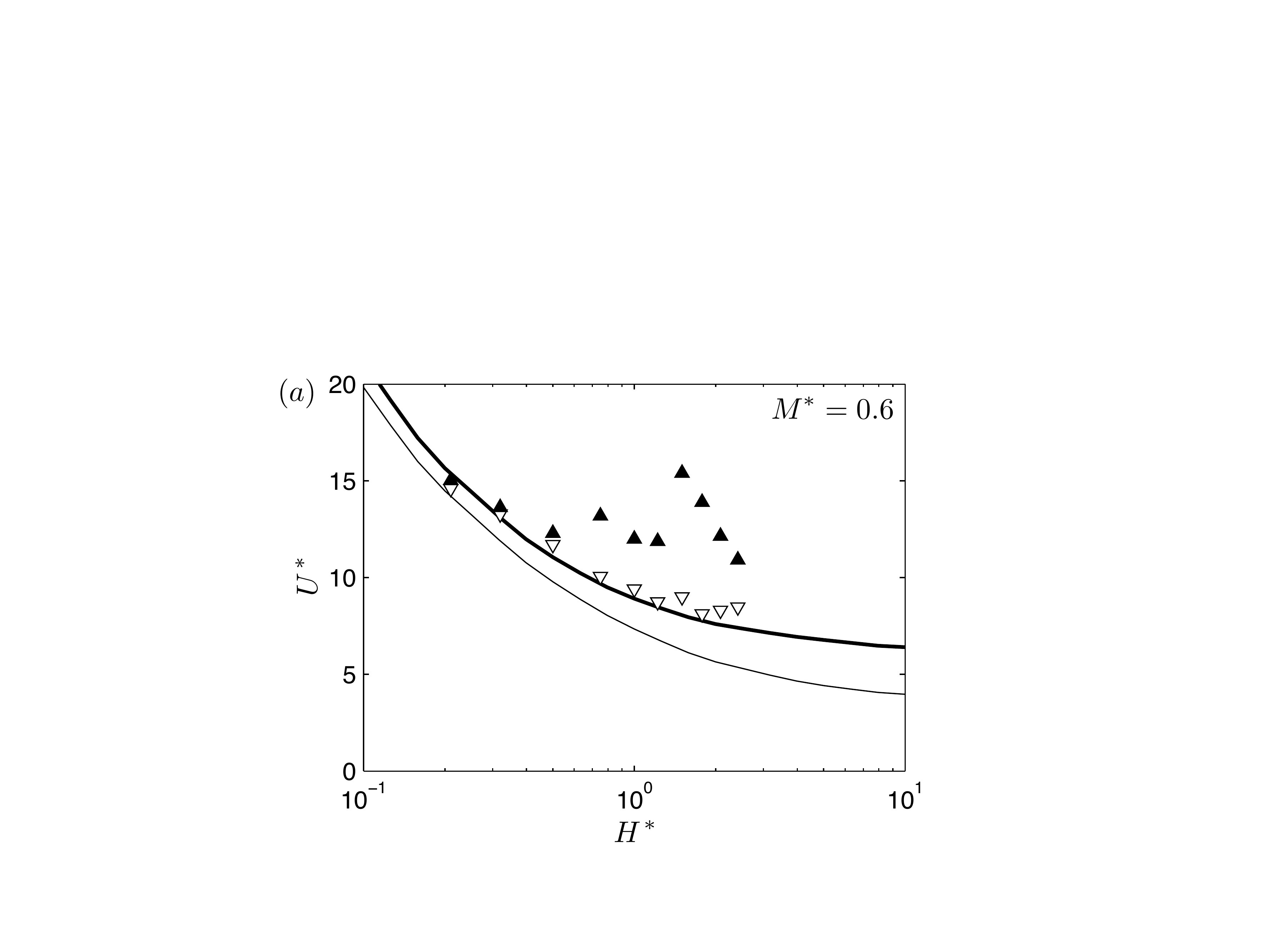}\includegraphics[scale=0.37]{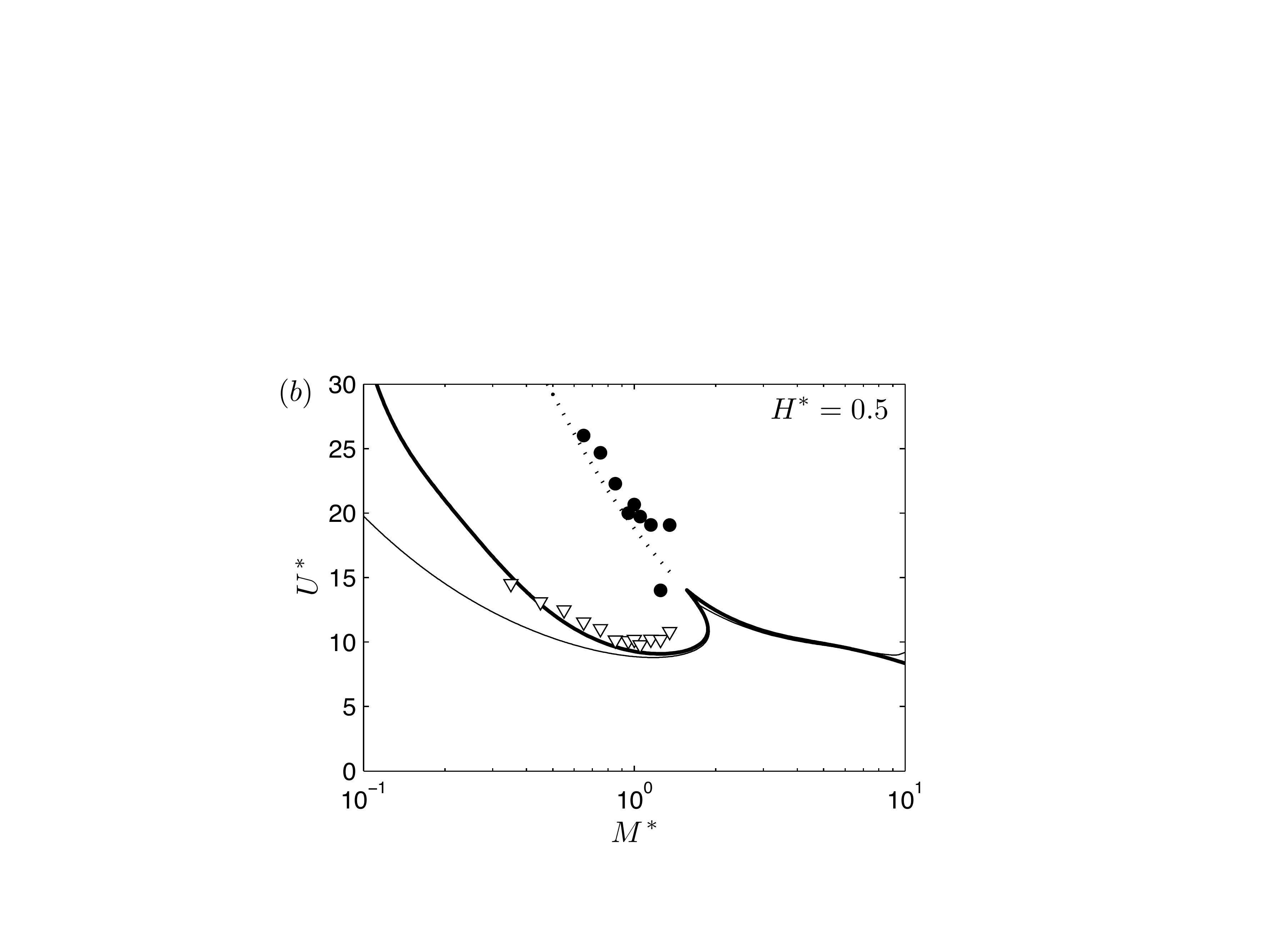}}
  \caption{Instability threshold as a function of $H^*$ for fixed $M^*=0.6$ (\textit{a}) and as a function of $M^*$ for fixed $H^*=0.5$ (\textit{b}) \textmd{for flat plates}. Open and filled triangles correspond to $U^*_d$ and $U^*_c$ respectively. The thick line is the result of the linear stability analysis detailed in \S\ref{sec:linear.model} and, the thin line is the same analysis without dissipative terms (i.e. for $\nu=\mu=0$). The filled circles in (\textit{b}) mark the departure from periodic motion  and the dotted line indicates the location where the second and third eigenmodes have the same frequency.}
\label{fig:UcvsH}
\end{figure}

Figure~\ref{fig:UcvsH} shows how the thresholds $U_d^*$ and $U^*_c$ vary with aspect ratio for constant mass ratio, or vary with mass ratio for constant aspect ratio. This measurement is compared with the predictions of the linear stability analysis and the agreement is excellent for $U_d^*$, confirming the proposed scenario. It also shows that taking into account the internal and viscous damping gives a better prediction of the instability threshold.

Figure~\ref{fig:UcvsH}\textit{b} also shows that, for sufficiently large airflow velocities, the flutter is no longer periodic. This departure from periodic motion occurs apparently when the unstable mode (with wavenumber $k^*\approx 3\pi/2$) has the same frequency as the next eigenmode (with $k^*\approx 5\pi/2$).
A nonlinear interaction between these modes is thus likely to be the cause of this secondary bifurcation, which is currently being investigated.


\section{Conclusion}
\label{sec:discussion}

In this paper, we addressed theoretically and experimentally the nonlinear dynamics of the flag instability near its threshold.
Weakly nonlinear analyses have been carried out both in the slender body limit and in the two-dimensional limit. In the former case, the bifurcation is always supercritical, while it can be subcritical in the latter case for large enough mass ratios \citep[in agreement with  numerical simulations of][]{Tang2007,Alben2008,Michelin2008}.

In the experiments, the dynamics near threshold is complicated by inherent planeity defects, which usually lead to large hysteresis.
These defects are \textmd{likely to be} the main cause of discrepancies between numerical simulations and experiments.
They are also responsible for the poor repeatability of the measured instability threshold when airflow velocity is increased. However, the threshold measured for decreasing velocity is both repeatable and in excellent agreement with the predictions of the linear stability analysis.

\begin{acknowledgments}
We warmly thank S\'ebastien Michelin for helpful discussions.
This study was partially funded by the French ANR (ANR-06-JCJC-0087).
C.~E. also acknowledges support from the European Commission (PIOF-GA-2009-252542).
\end{acknowledgments}

\bibliographystyle{jfm}

\bibliography{/Users/Ch/Documents/LaTeX/biblio}

\end{document}